\documentclass[a4paper,11pt]{article}

\usepackage{amsmath, amsfonts, amscd, amssymb}
\usepackage{amsthm}

\renewcommand{\baselinestretch}{1.5}

\begin{document}

\renewcommand{\baselinestretch}{1.5}

\vbox{\vspace{6mm}}
\begin{center}{ {\large \bf  TOPOS-THEORETIC RELATIVIZATION OF PHYSICAL REPRESENTABILITY AND QUANTUM GRAVITY}\\[7mm] \bf Elias Zafiris and Anastasios Mallios\\ {\it
University of Athens
\\ Department of Mathematics
\\ Panepistimioupolis, 15784 Athens
\\ Greece  \\} \vspace{2mm} }\end{center} \vspace{8mm}

\footnotetext{{\it E-mail }:{ \bf ezafiris@math.uoa.gr,
amallios@math.uoa.gr}}

\begin{abstract}
In the current debate referring to the construction of a tenable
background independent theory of Quantum Gravity we introduce the
notion of topos-theoretic relativization of physical
representability and demonstrate its relevance  concerning the
merging of General Relativity and Quantum Theory. For this purpose
we show explicitly that the dynamical mechanism of physical fields
can be constructed by purely algebraic means, in terms of
connection inducing functors and their associated curvatures, independently of any background substratum.
The application of this mechanism in General Relativity is
constrained by the absolute representability of the theory in the
field of real numbers. The relativization of physical
representability inside operationally selected topoi of sheaves
forces an appropriate interpretation of the mechanism of connection functors 
in terms of a generalized differential geometric dynamics of the corresponding
fields in the regime of these topoi. In particular, the relativization inside the topos of sheaves over commutative
algebraic contexts makes possible the formulation of quantum
gravitational dynamics by suitably adapting the functorial
mechanism of connections inside that topos.

\end{abstract}

\newpage

\renewcommand{\baselinestretch}{1.3}

\vbox{\vspace{6mm}}

\section{Prologue}
There exists a significant amount of current research in
theoretical physics devoted to the construction of a tenable
quantum theory of gravity, conceived as an extensive unifying
framework of both General Relativity and Quantum Theory [1-11]. It
has been generally argued that these fundamental physical theories
are based on incompatible conceptual and mathematical foundations.
In this sense, the task of their reconciliation in a unifying
framework, that respects the constraints posed by both theories,
requires a radical revision, or at least, a careful rethinking of
our current understanding of the basic notions, such as the
conception of spacetime, physical fields, localization,
observables and dynamics.

In this communication we would like to draw attention regarding
these issues from the algebraic, categorical and topos-theoretic perspective
of modern mathematics [12-19], as a substitute of the
set-theoretic one, and especially the impact of the consequences
of this perspective, in relation to their foundational
significance towards the crystallization of the basic notions
constituting a theory of Quantum Gravity. An initial motivation
regarding the relevance of the categorical viewpoint originates
from the realization that both of our fundamental theories can be
characterized in general terms as special instances of the
replacement of the constant by the variable. The semantics of this
transition, for both General Relativity and Quantum Theory, may be
incorporated in an algebraic topos-theoretic framework, that
hopefully provides the crucial pointers for the schematism of the
essential concepts needed for the intelligibilty of a theory of
Quantum Gravity, which respects the normative requirements of its
predecessors. Epigrammatically it is instructive to remark that,
in the case of General Relativity this process takes place through
the rejection of the fixed kinematical structure of spacetime, by
making the metric into a dynamical object determined solely by the
solution of Einstein's field equations. In the case of Quantum
Theory, the process of replacement of the constant by the
variable, is signified by the imposition of Heisenberg's
uncertainty relations, that determines the limits for simultaneous
measurements of certain pairs of complementary physical
properties, like position and momentum. Although this process in
Quantum Theory is not immediately transparent as in General
Relativity, it will eventually become clear that it is
indispensable to a unifying perspective.

From a mathematical point of view,
the general process of semantic transition from constant to variable structures 
is being effectuated by passing to appropriate topoi, where, an abstract topos
is conceived as a universe of variable sets, whose variation is being considered over
generalized localization domains. Thus, there exists the possibility of comprehending uniformly
the difference in the distinct instances of replacement of the constant by the variable, as they are
explicated in the concrete cases of General Relativity and Quantum Theory respectively, by
employing different topoi, corresponding to the localization properties of observables in
each theory. Of course, this strategy would be fruitful in a unifying quantum relativistic perspective, if we managed to disassociate the
dependence of dynamics in the regime of each theory from any fixed background spatiotemporal reference. Equivalently stated, the
dynamical mechanism should be ideally formulated functorially and purely algebraically. The benefit of such a formulation has to do with
the fact that, because of its functoriality, it can be algebraically forced uniformly inside the appropriate localization topoi of the above theories.
Thus, both of these theories can be treated homogenously regarding their dynamical mechanism, whereas, their difference can be traced to the
distinctive localization topoi employed in each case. In particular, the functorial representation of general relativistic gravitational dynamics induces a reformulation of the issue of quantization as a problem of selection of an appropriate localization topos, in accordance with the behaviour of observables in that regime, that effectuates the difference in the semantic interpretation of the dynamical machinery corresponding to the transition from the classical to the quantum case. In this work, we initially show that such a functorial dynamical mechanism can be actually constructed using methods of categorical homological algebra. More precisely, the homological dynamical mechanism is based on the modelling of the
notion of physical fields in terms of connections, which, effectuate
the functorial algebraic process of infinitesimal scalars extensions, due to
interactions caused by these fields. Subsequently, we explain the applicability of homological functorial dynamics to the problem of quantum gravity, according to the preceding remarks, by implementing the principle of topos-theoretic relativization of physical representability, using the technique of sheafification over appropriate localization domains.

The central focus of the categorical way of rethinking basic
notions in this endeavour can be described as a shift in the
emphasis of what is considered to be fundamental for the formation
of structures. In the set-theoretic mode of thinking, structures
of any conceivable form are defined as sets of elements endowed
with appropriate relations. In the category-theoretic mode, the
emphasis is placed on the transformations among the objects of a
category devised to represent a certain structure by means of
appropriate structural constraints on the collection of these
transformations. In this sense, the notion of structure does not
refer exclusively to a fixed universe of sets of predetermined
elements, but acquires a variable reference. We will argue that
this is an appealing feature, pertaining decisively to a revised
conceptualization of the basic notions, as above, in a viable
Quantum Gravity framework.

\section{A Homological Schema of Functorial Field Dynamics}
The basic conceptual and technical issue pertaining to the current research
attempts towards the construction of a tenable Quantum Gravity theory, refers to
the problem of independence of this theory from a fixed spacetime manifold substratum.
In relation to this problem, we demonstrate the existence and functionality of a homological schema of modelling general relativistic dynamics functorially, constructed by means of
connection inducing functors and their associated curvatures, which is, remarkably, independent of any background substratum.

\subsection{Algebraic Dynamicalization and Representability}

The basic defining feature of General Relativity, in
contradistinction to Newtonian classical theory, as well as
Special Relativity, is the abolishment of any fixed preexisting
kinematical framework by means of dynamicalization of the metric
tensor. This essentially means that, the geometrical relations
defined on a four dimensional manifold, making it into a
spacetime, become variable. Moreover, they are constituted
dynamically by the gravitation field, as well as other fields from
which matter can be derived, by means of Einstein's field
equations, through the imposition of a compatibility requirement
relating the metric tensor, which represents the spacetime geometry,
with the affine connection, which represents the gravitational
field. The dynamic variability of the geometrical structure on the
spacetime manifold constitutes the means of dynamicalization of
geometry in the descriptive terms of General Relativity,
formulated in terms of the differential geometric framework on
smooth manifolds. The intelligibility of the framework is enriched
by the imposition of the principle of general covariance of the
field equations under arbitrary coordinate transformations of the
points of the manifold preserving the differential structure,
identified as the group of manifold diffeomorphisms. As an
immediate consequence, the points of the manifold lose any
intrinsic physical meaning, in the sense that, they are not
dynamically localizable entities in the theory. Most importantly,
manifold points assume an indirect reference as indicators of
spacetime events only after the dynamical specification of
geometrical relations among them, as particular solutions of the
generally covariant field equations. From an algebraic viewpoint
[12-14, 21], a real differential manifold $M$ can be recovered completely
from the $\mathcal R$-algebra $\mathcal {C^\infty}({M})$ of smooth
real-valued functions on it, and in particular, the points of $M$
may be recovered from the algebra $\mathcal {C^\infty}({M})$ as
the algebra morphisms $\mathcal {C^\infty}({M}) \rightarrow
\mathcal R$.

In this sense, manifold points constitute the $\mathcal
R$-spectrum of $\mathcal {C^\infty}({M})$, being isomorphic with
the maximal ideals of that algebra. Notice that, the $\mathcal
R$-algebra $\mathcal {C^\infty}({M})$ is a commutative algebra that
contains the field of real numbers $\mathcal R$ as a distinguished
subalgebra. This particular specification incorporates the physical
assumption that our form of observation is being represented
globally by evaluations in the field of real numbers. In the
setting of General Relativity the form of observation is being
coordinatized by means of a commutative unital algebra of scalar
coefficients, called an algebra of observables, identified as the
$\mathcal R$-algebra of smooth real-valued functions $\mathcal
{C^\infty}({M})$. Hence, the background substratum of the theory
remains fixed as the $\mathcal R$-spectrum of the coefficient algebra
of scalars of that theory, and consequently, the points of the
manifold $M$, although not dynamically localizable degrees of
freedom of General Relativity, are precisely the semantic
information carriers of an absolute representability principle,
formulated in terms of global evaluations of the algebra of scalars
in the field of real numbers. Of course, at the level of the
$\mathcal R$-spectrum of $\mathcal {C^\infty}({M})$, the only
observables are the smooth functions evaluated over the points of
$M$. In physical terminology, the introduction of new observables
is conceived as the result of interactions caused by the presence
of a physical field, identified with the gravitational field in
the context of General Relativity. Algebraically, the process of
extending the form of observation with respect to the algebra of
scalars we have started with, that is $\mathcal A= \mathcal
{C^\infty}({M})$, due to field interactions, is described by means
of a fibering, defined as an injective morphism of $\mathcal
R$-algebras $\iota: \mathcal A \hookrightarrow \mathcal B$. Thus,
the $\mathcal R$-algebra $\mathcal B$ is considered as a module
over the algebra $\mathcal A$. A section of the fibering $\iota:
\mathcal A \hookrightarrow \mathcal B$, is represented by a
morphism of $\mathcal R$-algebras $s: \mathcal B \rightarrow
\mathcal A$, left inverse to $\iota$, that is $s \circ \iota=id_A$. The
fundamental extension of scalars of the $\mathcal R$-algebra
$\mathcal A$ is obtained by tensoring $\mathcal A$ with itself
over the distinguished subalgebra of the reals, that is $\iota:
\mathcal A \hookrightarrow \mathcal A {\bigotimes}_{\mathcal R}
\mathcal A$. Trivial cases of scalars extensions, in fact isomorphic
to $\mathcal A$, induced by the fundamental one, are obtained by tensoring $\mathcal A$ with
$\mathcal R$ from both sides, that is $\iota_1: \mathcal A
\hookrightarrow \mathcal A {\bigotimes}_{\mathcal R} \mathcal R$,
$\iota_2: \mathcal A \hookrightarrow \mathcal R
{\bigotimes}_{\mathcal R} \mathcal A$.

The basic idea of Riemann that has been incorporated in the
context of General Relativity is that geometry should be built
from the infinitesimal to the global. Geometry in this context is
understood in terms of metric structures that can be defined on a
differential manifold. If we adopt the algebraic viewpoint,
geometry as a result of interactions, requires the extension of
scalars of the algebra $\mathcal A$ by infinitesimal quantities,
defined as a fibration: $$d_* : \mathcal A \hookrightarrow
\mathcal A \oplus \mathbf M \cdot \epsilon$$ $$f \mapsto f + d_* (f)
\cdot \epsilon$$ where, $d_* (f)=:df$ is considered as the
infinitesimal part of the extended scalar, and $\epsilon$ the
infinitesimal unit obeying $\epsilon^2=0$ [20]. The algebra of
infinitesimally extended scalars $\mathcal A \oplus \mathbf M
\cdot \epsilon$ is called the algebra of dual numbers over $\mathcal
A$ with coefficients in the $\mathcal A$-module $\mathbf M$. It is
immediate to see that the algebra $\mathcal A \oplus \mathbf M \cdot
\epsilon$, as an abelian group is just the direct sum $\mathcal A
\oplus \mathbf M$, whereas the multiplication is defined by: $$(f+
df \cdot \epsilon) \bullet (f^{'} + {df}^{'} \cdot \epsilon)=f
\cdot f^{'} + (f \cdot {df}^{'} +  f^{'} \cdot df) \cdot
\epsilon$$ Note that, we further require that the composition of
the augmentation $\mathcal A \oplus \mathbf M \cdot \epsilon
\rightarrow \mathcal A$, with $d_*$ is the identity. Equivalently,
the above fibration, viz., the homomorphism of algebras $d_* :
\mathcal A \hookrightarrow \mathcal A \oplus \mathbf M \cdot
\epsilon$, can be formulated as a derivation, that is, in terms of
an additive $\mathcal R$-linear morphism: $$d : \mathcal A
\rightarrow \mathbf M$$ $$f \mapsto df$$ that, moreover, satisfies
the Leibniz rule: $$d(f \cdot g)= f \cdot dg + g \cdot df$$ Since
the formal symbols of differentials $\{df, f \in \mathcal A\}$,
are reserved for the universal derivation, the $\mathcal A$-module
$\mathbf M$ is identified as the free $\mathcal A$-module $\mathbf
{\Omega}$ of $1$-forms generated by these formal symbols, modulo
the Leibniz constraint, where the scalars of the distinguished
subalgebra $\mathcal R$, that is the real numbers, are treated as
constants. K\"{a}hler observed that the free $\mathcal A$-module
$\mathbf \Omega$ can be constructed explicitly from the
fundamental form of scalars extension of $\mathcal A$, that is
$\iota: \mathcal A \hookrightarrow \mathcal A
{\bigotimes}_{\mathcal R} \mathcal A$ by considering the morphism:
$$\delta: \mathcal A {\bigotimes}_{\mathcal R} \mathcal A
\rightarrow A$$ $$f_1 \otimes f_2 \mapsto f_1 \cdot f_2$$ Then by
taking the kernel of this morphism of algebras, that is, the ideal:
$$\mathbf I= \{f_1 \otimes f_2 \in \mathcal A
{\bigotimes}_{\mathcal R} \mathcal A : \delta (f_1 \otimes f_2)=0
\} \subset \mathcal A {\bigotimes}_{\mathcal R} \mathcal A$$ it
can be shown that the morphism of $\mathcal A$-modules:
$$\Sigma : \mathbf \Omega \rightarrow \frac{\mathbf I}{{\mathbf
I}^2}$$ $$df \mapsto 1 \otimes f - f \otimes 1$$ is an isomorphism.

We can prove the above isomorphism as follows: The fractional object $\frac{\mathbf I}{{\mathbf
I}^2}$ has an $\mathcal A$-module structure defined by: $$f \cdot (f_1 \otimes f_2)= (f \cdot f_1) \otimes f_2= f_1 \otimes (f \cdot f_2)$$ for $f_1 \otimes f_2$ $\in$ $\mathbf I$, $f$ $\in$ $\mathcal A$. We can check that the second equality is true by proving that the difference of $(f \cdot f_1) \otimes f_2$ and $f_1 \otimes (f \cdot f_2)$ belonging to $\mathbf I$, is actually an elememt of ${{\mathbf
I}^2}$, viz., the equality is true modulo ${{\mathbf
I}^2}$. So we have: $$(f \cdot f_1) \otimes f_2 - f_1 \otimes (f \cdot f_2)= (f_1 \otimes f_2) \cdot (f \otimes 1 - 1 \otimes f)$$ The first factor of the above product of elements belongs to $\mathbf I$ by assumption, whereas the second factor also belongs to $\mathbf I$, since we have that: $$\delta 
(f \otimes 1 - 1 \otimes f)= 0$$ Hence the product of elements above belongs to $\mathbf I \cdot \mathbf I= {\mathbf I}^2$. Consequently, we can define a morphism of $\mathcal A$-modules: $$\Sigma : \mathbf \Omega \rightarrow \frac{\mathbf I}{{\mathbf
I}^2}$$ $$df \mapsto 1 \otimes f - f \otimes 1$$  Now, we construct the inverse of that morphism as follows: The $\mathcal A$-module $\mathbf \Omega$ can be made an ideal in the algebra of dual numbers over $\mathcal
A$, viz.,  $\mathcal A \oplus \mathbf \Omega
\cdot \epsilon$. Moreover, we can define the morphism of algebras: $$\mathcal A \times \mathcal A \rightarrow \mathcal A \oplus \mathbf \Omega
\cdot \epsilon$$ $$(f_1, f_2) \mapsto f_1 \cdot f_2 + f_1 \cdot d f_2 \epsilon$$ This is an $\mathcal R$-bilinear morphism of algebras, and thus, it gives rise to a morphism of algebras: $$\Theta: \mathcal A {{\otimes}_{\mathcal R}} \mathcal A \rightarrow \mathcal A \oplus \mathbf \Omega
\cdot \epsilon$$ Then, by definition we have that $\Theta (\mathbf I) \subset \mathbf \Omega$, and also, $\Theta( {\mathbf I}^2 )=0$. Hence, there is obviously induced a morphism of $\mathcal A$-modules: $$\mathbf \Omega \leftarrow \frac{\mathbf I}{{\mathbf
I}^2}$$ which is the inverse of $\Sigma$. Consequently, we conclude that: $$\mathbf \Omega \cong \frac{\mathbf I}{{\mathbf
I}^2}$$

Thus the free $\mathcal A$-module $\mathbf \Omega$ of
$1$-forms is isomorphic with the free $\mathcal A$-module
$\frac{\mathbf I}{{\mathbf I}^2}$ of K\"{a}hler differentials of the
algebra of scalars $\mathcal A$ over $\mathcal R$, conceived as a
distinguished ideal in the algebra of infinitesimally extended
scalars $\mathcal A \oplus \mathbf \Omega \cdot \epsilon$, due to
interaction, according to the following split short exact
sequence: $$\mathbf \Omega \hookrightarrow \mathcal A \oplus
\mathbf \Omega \cdot \epsilon \twoheadrightarrow \mathcal A$$ or
equivalently formulated as: $$0 \rightarrow \mathbf {\Omega}_A
\rightarrow  \mathcal A {\bigotimes}_{\mathcal R} \mathcal A
\rightarrow \mathcal A$$ By dualizing, we obtain the dual
$\mathcal A$-module of $\mathbf \Omega$, that is $\mathbf \Xi : =
Hom(\mathbf \Omega, \mathcal A)$. Thus we have at our disposal,
expressed in terms of infinitesimal scalars extension of the algebra
$\mathcal A$, semantically intertwined with the generation of
geometry as a result of interaction, new types of observables
related with the incorporation of differentials and their duals,
called vectors. Let us now explain the functionality of geometry,
as related with the infinitesimally extended rings of scalars
defined above, in the context of General Relativity. As we have
argued before, the absolute representability principle of this
theory, necessitates that our form of observation is tautosemous
with real numbers representability. This means that all types of
observables should possess uniquely defined dual types of
observables, such that their representability can be made possible
my means of real numbers. This is exactly the role of a geometry
induced by a metric. Concretely, a metric structure assigns a
unique dual to each observable, by effectuating an isomorphism
between the $\mathcal A$-modules $\mathbf \Omega$ and $\mathbf
\Xi= Hom(\mathbf \Omega, \mathcal A)$, that is: $$g: \mathbf
\Omega \simeq \mathbf \Xi$$ $$df \mapsto v_f := g (df)$$ Thus the
functional role of a metric geometry forces the observation of
extended scalars, by means of representability in the field of
real numbers, and is reciprocally conceived as the result of
interactions causing infinitesimal variations in the scalars of
the $\mathcal R$-algebra $\mathcal A$.

Before proceeding further, it is instructive at this point to
clarify the meaning of a universal derivation, playing a
paradigmatic role in the construction of extended algebras of scalars
as above, in appropriate category-theoretic terms as follows [20]: The
covariant functor of left ${\mathcal A}$-modules valued
derivations of ${\mathcal A}$: $$\overleftarrow {\mathcal
{{\nabla}}}_{{\mathcal A}}(-): \mathcal M^{({\mathcal A})}
\rightarrow \mathcal M^{({\mathcal A})}$$ is being representable
by the left ${\mathcal A}$-module of $1$-forms ${\mathbf
{\Omega}^1({\mathcal A})}$ in the category of left ${\mathcal
A}$-modules $\mathcal M^{({\mathcal A})}$, according to the
isomorphism: $$\overleftarrow {\mathcal {{\nabla}}}_{{\mathcal
A}}(N)\cong {Hom}_{{\mathcal A}}({\mathbf {\Omega}}^1({\mathcal
A}),N)$$ Thus, ${\mathbf {\Omega}^1({\mathcal A})}$ is
characterized categorically as a universal object in $\mathcal
M^{({\mathcal A})}$, and the derivation: $$d: {\mathcal A}
\rightarrow {\mathbf {\Omega}^1({\mathcal A})}$$ as the universal
derivation [20]. Furthermore, we can define algebraically, for
each $n \in N$, $n \geq 2$, the $n$-fold exterior product:
$${{\mathbf {{\mathbf {\Omega}}}}^n({\mathcal A})}={{\bigwedge}^n}
{{\mathbf {{\mathbf {\Omega}}}}^1({\mathcal A})}$$ where ${\mathbf
{{\mathbf {\Omega}}}}({\mathcal A}):={{\mathbf {{\mathbf
{\Omega}}}}^1({\mathcal A})}$, ${\mathcal A}:={{\mathbf {{\mathbf
{\Omega}^0}}}({\mathcal A})}$, and finally show analogously that
the left ${\mathcal A}$-modules of $n$-forms ${\mathbf {\Omega}}^n
({\mathcal A})$ in $\mathcal M^{({\mathcal A})}$ are representable
objects in $\mathcal M^{({\mathcal A})}$ of the covariant functor
of left ${\mathcal A}$-modules valued $n$-derivations of
${\mathcal A}$, denoted by ${{\overleftarrow {\mathbf
{\nabla}}}^n}_{{\mathcal A}}(-): \mathcal M^{({\mathcal A})}
\rightarrow \mathcal M^{({\mathcal A})}$. We conclude that, all
infinitesimally extended algebras of scalars that have been
constructed from ${\mathcal A}$ by fibrations, presented
equivalently as derivations, are representable as left ${\mathcal
A}$-modules of $n$-forms ${\mathbf {\Omega}}^n ({\mathcal A})$ in
the category of left ${\mathcal A}$-modules  $\mathcal
M^{({\mathcal A})}$.

We emphasize that the intelligibility of the algebraic schema is
based on the conception that infinitesimal variations in the
scalars of ${\mathcal A}$, are caused by interactions, meaning
that they are being effectuated by the presence of a physical
field, identified as the gravitational field in the context of
General Relativity. Thus, it is necessary to establish a purely
algebraic representation of the notion of physical field and
explain the functional role it assumes for the interpretation of
the theory. The key idea for this purpose amounts to expressing
the process of scalars extension in functorial terms, and by
anticipation identify the functor of infinitesimal scalars
extension due to interaction with the physical field that causes
it [20]. Regarding the first step of this strategy we clarify that the
general process of scalars extension from an algebra ${\mathcal S}$ to
an algebra ${\mathcal T}$ is represented functorially by means of the
functor of scalars extension [12], from ${\mathcal S}$ to
${\mathcal T}$ as follows: $$\mathbf F: \mathcal M^{({\mathcal
S})} \rightarrow \mathcal M^{({\mathcal T})}$$ $$\mathbf E \mapsto
\mathbf T {\bigotimes}_{{\mathcal S}} \mathbf E $$ The second step
involves the application of the functorial algebraic procedure for
the case admitting the identifications: $${\mathcal S}={\mathcal
A}$$ $${\mathcal T}=[\mathcal A \oplus {\mathbf
{\Omega}^1({\mathcal A})} \cdot \epsilon]$$ corresponding to
infinitesimal scalars extension. Consequently, the physical field
as the causal agent of interactions admits a purely algebraic
description as the functor of infinitesimal scalars extension,
called a connection-inducing functor: $$\widehat{{\mathcal
{{\nabla}}}}: \mathcal M^{({\mathcal A})} \rightarrow \mathcal
M^{({{\mathcal A \oplus {\mathbf {\Omega}^1({\mathcal A})}}}\cdot
\epsilon)}$$ $$\mathbf E \mapsto [\mathcal A \oplus {\mathbf
{\Omega}^1({\mathcal A})} \cdot \epsilon ] {\bigotimes}_{{\mathcal
A}} \mathbf E $$ In this sense, the vectors of the left ${\mathcal
A}$-module $\mathbf E$ are being infinitesimally extended into
vectors of the left ${({{\mathcal A \oplus {\mathbf
{\Omega}^1({\mathcal A})}}}\cdot \epsilon)}$-module $[\mathcal A
\oplus {\mathbf {\Omega}^1({\mathcal A})} \cdot \epsilon ]
{\bigotimes}_{{\mathcal A}} \mathbf E $. Notice that these two
kinds of vectors are being defined over different algebras. Hence, in
order to compare them we have to pull the infinitesimally extended
ones back to the initial algebra of scalars, viz., the $\mathcal
R$-algebra $\mathcal A$. Algebraically this process is implemented
by restricting the left ${({{\mathcal A \oplus {\mathbf
{\Omega}^1({\mathcal A})}}}\cdot \epsilon)}$-module $[\mathcal A
\oplus {\mathbf {\Omega}^1({\mathcal A})} \cdot \epsilon ]
{\bigotimes}_{{\mathcal A}} \mathbf E $ to the $\mathcal
R$-algebra $\mathcal A$. If we perform this base change we obtain
the left ${\mathcal A}$-module $\mathbf E \oplus [{\mathbf
{\Omega}^1({\mathcal A})} {\bigotimes}_{{\mathcal A}} \mathbf E]
\cdot \epsilon $. Thus, the effect of the action of the physical
field on the vectors of the left ${\mathcal A}$-module $\mathbf E$
can be expressed by means of the following comparison morphism of
left ${\mathcal A}$-modules: $${{{\mathcal
{{\nabla}}}^\star}}_{\mathbf E}: \mathbf E \rightarrow \mathbf E
\oplus [{\mathbf {\Omega}^1({\mathcal A})} {\bigotimes}_{{\mathcal
A}} \mathbf E] \cdot \epsilon $$ Equivalently, the irreducible amount of information
incorporated in the comparison morphism can be now expressed as a
connection on $\mathbf E$. The latter is defined algebraically as an $\mathcal R$-linear
morphism of ${\mathcal A}$-modules [21]: $${{\mathcal
{{\nabla}}}}_{\mathbf E}: \mathbf E \rightarrow {\mathbf
{\Omega}^1({\mathcal A})} {\bigotimes}_{{\mathcal A}} \mathbf E=
\mathbf E {\bigotimes}_{{\mathcal A}} {\mathbf
{\Omega}^1({\mathcal A})}={\mathbf {\Omega}^1({\mathbf E})}$$ such
that the following Leibniz type constraint is satisfied:
$${{\mathcal {{\nabla}}}}_{\mathbf E} (f \cdot v)= f \cdot
{{\mathcal {{\nabla}}}}_{\mathbf E} (v) + df \otimes v$$ for all
$f \in \mathcal A$, $v \in \mathbf E$.

In the context of General Relativity, the absolute
representability principle over the field of real numbers,
necessitates as we have explained above the existence of uniquely
defined duals of observables. Thus, the gravitational field is
identified with a linear connection on the $\mathcal A$-module
$\mathbf \Xi= Hom({\mathbf \Omega}^1, \mathcal A)$, being isomorphic
with ${\mathbf \Omega}^1$, by means of a metric: $$g: {\mathbf
\Omega}^1 \simeq \mathbf \Xi= {{\mathbf \Omega}^1}^*$$
Consequently, the gravitational field may be represented by the
pair $(\Xi, {{\mathcal {{\nabla}}}}_{\mathbf \Xi})$. The metric
compatibility of the connection required by the theory is simply
expressed as: $${{\mathcal {{\nabla}}}}_{{{Hom}_{\mathcal A}}
{(\mathbf \Xi, {\mathbf \Xi}^*)}} (g)=0$$

It is instructive to emphasize that the functorial conception of
physical fields in general, according to the proposed schema,
based on the notion of causal agents of infinitesimal scalars
extension, does not depend on any restrictive representability
principle, like the absolute representability principle over the
real numbers, imposed by General Relativity. Consequently, the
meaning of functoriality implies covariance with respect to
representability, and thus, covariance with respect to generalized
geometric realizations. In the same vein of ideas, the reader has
already noticed that all the algebraic arguments refer, on
purpose, to a general observables algebra $\mathcal A$, that has been
identified with the $\mathcal R$-algebra $\mathcal
{C^\infty}({M})$ in the model case of General Relativity. Of
course, the functorial mechanism of understanding the notion of
interaction, should not depend on the observables algebras used
for the particular manifestations of it, thus, the only actual
requirement for the intelligibility of functoriality of
interactions by means of physical fields rests on the
algebra-theoretic specification of what we characterize structures of
observables. Put differently, the functorial coordinatization of
the universal mechanism of encoding physical interactions in terms
of observables, by means of causal agents, namely physical fields
effectuating infinitesimal scalars extension, should respect the
algebra-theoretic structure.

\subsection{The Algebraic De Rham Complex and Field Equations}

The next stage of development of the algebraic schema of
comprehending the mechanism of dynamics involves the satisfaction
of appropriate global constraints, that impose consistency
requirements referring to the transition from the infinitesimal to
the global. For this purpose it is necessary to employ the
methodology of homological algebra. We start by reminding the
algebraic construction, for each $n \in N$, $n \geq 2$, of the
$n$-fold exterior product as follows: ${{\mathbf {{\mathbf
{\Omega}}}}^n({\mathcal A})}={{\bigwedge}^n} {{\mathbf {{\mathbf
{\Omega}}}}^1({\mathcal A})}$ where ${\mathbf {{\mathbf
{\Omega}}}}({\mathcal A}):={{\mathbf {{\mathbf
{\Omega}}}}^1({\mathcal A})}$, ${\mathcal A}:={{\mathbf {{\mathbf
{\Omega}^0}}}({\mathcal A})}$. We notice that there exists an
$\mathcal R$-linear morphism: $$d^n : {{\mathbf {{\mathbf
{\Omega}}}}^n({\mathcal A})} \rightarrow {{\mathbf {{\mathbf
{\Omega}}}}^{n+1} ({\mathcal A})}$$ for all $n \geq 0$, such that
$d^0=d$. Let $\omega \in {{\mathbf {{\mathbf
{\Omega}}}}^n({\mathcal A})}$, then $\omega$ has the form:
$$\omega= \sum {f_i (dl_{i1} \bigwedge {\ldots} \bigwedge
dl_{in})}$$ with $f_i$, $l_{ij}$, $\in$ $\mathcal A$ for all
integers $i$, $j$. Further, we define: $$d^n (\omega)= \sum {df_i
\bigwedge dl_{i1} \bigwedge {\ldots} \bigwedge dl_{in}}$$ Then, we
can easily see that the resulting sequence of $\mathcal R$-linear
morphisms; $$\mathcal A \rightarrow {{\mathbf {{\mathbf
{\Omega}}}}^1({\mathcal A})} \rightarrow {\ldots} \rightarrow
{{\mathbf {{\mathbf {\Omega}}}}^n({\mathcal A})} \rightarrow
{\ldots}$$  is a complex of $\mathcal R$-vector spaces, called the
algebraic de Rham complex of $\mathcal A$. The notion of complex
means that the composition of two consequtive $\mathcal R$-linear
morphisms vanishes, that is $d^{n+1} \circ d^n=0$, simplified
symbolically as: $$d^{\mathbf 2}=0$$ If we assume that $(\mathbf
E, {{\mathcal {{\nabla}}}}_{\mathbf E})$ is an interaction field,
defined by a connection ${{\mathcal {{\nabla}}}}_{\mathbf E}$ on
the $\mathcal A$-module $\mathbf E$, then ${{\mathcal
{{\nabla}}}}_{\mathbf E}$ induces a sequence of $\mathcal
R$-linear morphisms: $$\mathbf E \rightarrow {{\mathbf {{\mathbf
{\Omega}}}}^1({\mathcal A})} {\bigotimes}_{{\mathcal A}} \mathbf E
\rightarrow {\ldots} \rightarrow {{\mathbf {{\mathbf
{\Omega}}}}^n({\mathcal A})} {\bigotimes}_{{\mathcal A}} \mathbf E
\rightarrow {\ldots}$$ or equivalently: $$\mathbf E \rightarrow
{{\mathbf {{\mathbf {\Omega}}}}^1({\mathbf E})}  \rightarrow
{\ldots} \rightarrow {{\mathbf {{\mathbf {\Omega}}}}^n({\mathbf E
})} \rightarrow {\ldots}$$ where the morphism: $${{\mathcal
{{\nabla}}}}^n : {{\mathbf {{\mathbf {\Omega}}}}^n({\mathcal A})}
{\bigotimes}_{{\mathcal A}} \mathbf E \rightarrow {{\mathbf
{{\mathbf {\Omega}}}}^{n+1}({\mathcal A})} {\bigotimes}_{{\mathcal
A}} \mathbf E$$ is given by the formula: $${{\mathcal
{{\nabla}}}}^n(\omega \otimes v)= d^n (\omega) \otimes v + (-1)^n
\omega \wedge {{\mathcal {{\nabla}}}}(v)$$ for all $\omega$ $\in$
${{\mathbf {{\mathbf {\Omega}}}}^n({\mathcal A})}$, $v$ $\in$
$\mathbf E$. It is immediate to see that ${{\mathcal
{{\nabla}}}}^0={{\mathcal {{\nabla}}}_\mathbf E}$. Let us denote
by: $${\mathbf R}_\nabla : \mathbf E \rightarrow {{\mathbf
{{\mathbf {\Omega}}}}^2({\mathcal A})} {\bigotimes}_{{\mathcal A}}
\mathbf E= {{\mathbf {{\mathbf {\Omega}}}}^2({\mathbf E})}$$ the
composition ${{\mathcal {{\nabla}}}}^1 \circ {{\mathcal
{{\nabla}}}}^0$. We see that ${\mathbf R}_\nabla$ is actually an
$\mathcal A$-linear morphism, that is $\mathcal A$-covariant, and
is called the curvature of the connection ${{\mathcal
{{\nabla}}}_\mathbf E}$. We note that, for the case of the
gravitational field $(\Xi, {{\mathcal {{\nabla}}}}_{\mathbf
\Xi})$, in the context of General Relativity, ${\mathbf R}_\nabla$
is tautosemous with the Riemannian curvature of the spacetime
manifold. We notice that, the latter sequence of $\mathcal
R$-linear morphisms, is actually a complex of $\mathcal R$-vector
spaces if and only if: $${\mathbf R}_\nabla=0$$ We say that the
connection ${{\mathcal {{\nabla}}}_\mathbf E}$ is integrable if
${\mathbf R}_\nabla=0$, and we refer to the above complex as the
de Rham complex of the integrable connection ${{\mathcal
{{\nabla}}}_\mathbf E}$ on $\mathbf E$ in that case. It is also
usual to call a connection ${{\mathcal {{\nabla}}}_\mathbf E}$
flat if ${\mathbf R}_\nabla=0$. A flat connection defines a
maximally undisturbed process of dynamical variation caused by the
corresponding physical field. In this sense, a non-vanishing
curvature signifies the existence of disturbances from the
maximally symmetric state of that variation. These disturbances
can be cohomologically identified as obstructions to deformation
caused by physical sources. In that case, the algebraic de Rham
complex of the algebra of scalars $\mathcal A$ is not acyclic, viz.
it has non-trivial cohomology groups. These groups measure the
obstructions caused by sources and are responsible for a
non-vanishing curvature of the connection. In the case of General
Relativity, these disturbances are associated with the presence of
matter distributions, being incorporated in the specification of
the energy-momentum tensor. Taking into account the requirement of
absolute representability over the real numbers, and thus
considering the relevant evaluation trace operator by means of the
metric, we arrive at Einstein's field equations, which in the
absence of matter sources read: $${\mathcal R}({{\mathcal
{{\nabla}}}_\mathbf \Xi})=0$$ where ${\mathcal R}({{\mathcal
{{\nabla}}}_\mathbf \Xi})$ denotes the relevant Ricci scalar
curvature.

\section{Topos-Theoretic Relativization and Localization of Functorial Dynamics}

\subsection{Conceptual Setting}
The central focus of the studies pertaining to the formulation of a
tenable theory of Quantum Gravity resolves around the issue of
background manifold independence. In Section 2,
we have constructed a general functorial framework of modelling field dynamics 
using categorical and homological algebraic concepts and techniques. In particular, we have 
applied this framework in the case of General Relativity recovering the classical
gravitational dynamics. The significance of the proposed functorial schema of dynamics, in relation to
a viable topos-theoretic approach to Quantum Gravity,
lies on the fact that, the coordinatization of
the universal mechanism of encoding physical interactions in terms
of observables, by means of causal agents, viz., physical fields
effectuating infinitesimal scalars extension, should respect only the
algebra-theoretic structure of observables. Consequently, it is not dependent on the
codomain of representability of the observables, being thus, only subordinate to the algebra-theoretic
characterization of their structures. In particular, it not constrained at all by
the absolute representability principle over the field of 
real numbers, imposed by classical General Relativity, a byproduct of which is the fixed background 
manifold construct of that theory.

In this perspective, the
absolute representability principle of classical General Relativity in terms
of real numbers, may be relativized without affecting the
functionality of the algebraic dynamical mechanism. Consequently, it is possible to describe the dynamics of
gravitational interactions in generalized localization environments, instantiated by suitable topoi. 
The latter are understood in the sense of categories of sheaves, defined over a base category of reference localization contexts, with
respect to some suitable Grothendieck topology. From a physical viewpoint, the construction of a sheaf of observables constitutes the 
natural outcome of a well-defined localization process. Generally speaking, a localization process is being
implemented in terms of an action of some category of reference contexts
on a set-theoretic global algebra of observables. The
latter, is then partitioned into sorts parameterized by the
objects of the category of contexts. In this manner, the 
functioning of a localization process can be represented by means
of a fibered construct, understood geometrically as a presheaf, or equivalently, as a variable
set (algebra) over the base category of contexts. The fibers of this
construct may be thought, in analogy to the case of the action of
a group on a set of points, as the ``generalized orbits" of the
action of the category of contexts.  The notion of functional
dependence incorporated in this action, forces the global algebraic
structure of observables to fiber over the base category of
reference contexts.

At a further stage of development of these ideas, the disassociation of the physical meaning of a localization process from its usual classical spatiotemporal connotation, 
requires, first of all, the abstraction of the constitutive properties of localization in appropriate categorical terms, and then, the
effectuation of these properties for the definition of localization systems of global observable structures. Regarding these objectives, the sought abstraction is being implemented by means of covering devices on the base category of
reference contexts, called in categorical terminology covering sieves. The constitutive properties of localization being abstracted categorically in terms of sieves, being qualified as covering ones, satisfy the following basic requirements:

[i]. The covering sieves are covariant under pullback
operations, viz., they are stable under change of a base reference context. Most importantly, the stability conditions are functorial. This requirement means, in particular, that the intersection of covering sieves is also a covering
sieve, for each reference context in the base category.

[ii]. The covering sieves are transitive, such
that, intuitively stated,  covering sieves of subcontexts of a context
in covering sieves of this context, are also covering sieves
of the context themselves.

From a physical perspective, the consideration of covering sieves
as generalized measures of localization of observables within a global observable structure, gives rise to localization systems of the latter. More specifically, the operation which assigns to each local reference context of the base category a collection of covering sieves satisfying the closure conditions stated previously, gives rise to the notion of a Grothendieck topology on the category of contexts. The construction of a suitable Grothendieck topology on the base category of contexts is  significant for the following reasons: Firstly, it elucidates precisely and unquestionably the conception of local in a categorical measurement environment, such that, this conception becomes detached from its usual spatiotemporal connotation, and thus, expressed exclusively in relational information terms. Secondly, it permits the collation of local observable information into global ones by utilization of the notion of sheaf for that Grothendieck topology. The definition of sheaf essentially expresses gluing conditions, providing the means for studying the global consequences of locally defined properties. The transition
from locally defined observable information into global ones
is being effectuated via a compatible family of elements over a localization system of a global observable strucure. A sheaf assigns a set of elements to each reference context of a localization system, representing local observable data colected within that context. A choice of
elements from these sets, one for each context, forms a compatible
family if the choice respects the mappings induced by the restriction
functions among contexts, and moreover, if the elements chosen agree whenever two contexts of
the localization system overlap. If such a locally compatible choice
induces a unique choice for a global observable structure being localized, viz. a
global choice, then the condition for being a sheaf is satisfied.
We note that, in general, there will be more locally defined or
partial choices than globally defined ones, since not all partial
choices need be extendible to global ones, but a compatible family
of partial choices uniquely extends to a global one, or in other
words, any presheaf uniquely defines a sheaf.

According to the strategy of relativization of physical representability with respect to generalized localization environments, instantiated by suitable categories of sheaves of observables, the problem of quantization of gravity is equivalent to forcing the algebraic general relativistic dynamical mechanism of the gravitational connection functorial morphism inside an appropriate topos, being capable of incorporating the localization properties of observables in the quantum regime. The only cost to be paid
for this topos-theoretic relativization is the rejection of the fixed background
manifold structure of the classical theory. This is actually not a cost at
all, since it would permit the intelligibility of the field
equations over geometric realizations that include manifold
singularities and other pathologies, without affecting the
algebraic mechanism of dynamics.

Equivalently stated, the requirement of background manifold
independence of Quantum Gravity can be attained, by rejecting the absolute representability of
the classical theory over the real numbers, and thus, the fixed spacetime manifold substratum, while keeping at
the same time, the homological machinery of functorial dynamics. In the current Section,
we argue that the abolishment of the above absolute representability requirement of the classical theory,
paving the way towards Quantum Relativity, 
can be achieved by effectuating a process of topos-theoretic relativization of physical 
representability suitable for the modelling of quantum phenomena. This process is based on
the technique of sheafification of observables algebras and constitutes the necessary conceptual and technical
apparatus for a merging of General Relativity and Quantum Theory along the topos-theoretic lines advocated in this work.

\subsection{Topological Sheafification of Field Dynamics and Abstract Differential Geometry}

Initially, it is important to clarify the
conception of relativity of representability in appropriate
category-theoretic terms.
The absolute representability principle is based on the
set-theoretic conception of the real line, as a set of infinitely
distinguished points coordinatized by means of the field of real
numbers. Expressed categorically, this is equivalent to the
interpretation of the algebraic structure of the reals inside the
absolute universe of $\mathbf {Sets}$, or more precisely inside
the topos of constant $\mathbf {Sets}$. It is also well known that
algebraic structures and mechanisms can admit a variable
reference, formulated in category-theoretic jargon in terms of
arrows only specifications, inside any suitable topos of discourse
[13, 15]. A general topos can be conceived as a manifestation of a
universe of variable sets [17, 18]. For example the topos of
sheaves of sets $\mathbf {Shv}(X)$ over the category of open sets
of an abstract topological space $X$, ordered by inclusion, is
understood as a categorical universe of varying sets over the
opens of the topology covering $X$. The relativization of physical
representability with respect to the topos of sheaves $\mathbf
{Shv}(X)$, amounts to the relativization of both the notion and
the algebraic structure of the real numbers inside this topos [22].
Regarding the notion of real numbers inside the topos $\mathbf
{Shv}(X)$, this is equivalent to the notion of continuously
variable real numbers over the open reference domains of $X$, or else, equivalent to
the notion of real-valued continuous functions on $X$, when
interpreted respectively inside the topos of $\mathbf {Sets}$ [17,
18] . Regarding the algebraic structure of the reals inside the
topos $\mathbf {Shv}(X)$, they form only an algebra in this topos,
which is tautosemous with the sheaf of commutative 
$\mathcal R$-algebras of continuous real-valued functions on $X$,
where $\mathcal R$ corresponds in that case to the constant sheaf
of real numbers over $X$.

Let us discuss briefly from a physical viewpoint, the meaning of
relativization of representability with respect to the internal
reals of the topos of sheaves $\mathbf {Shv}(X)$ of $\mathcal
R$-algebras of observables $\mathcal A$ over the category of opens
of $X$. Inside this topos, it is assumed that for every open $U$
in $X$, ${\mathcal A}(U)$ is a commutative, unital $\mathcal
R$-algebra of continuous local sections of the sheaf of $\mathcal
R$-algebras $\mathcal A$. In particular, the algebra of reals in this
topos consists of continuous real-valued local sections localized
sheaf-theoretically over the opens of $X$. Thus, the semantics of
the codomain of valuation of observables is transformed from a
set-theoretic to a sheaf-theoretic one. More concretely, it is
obvious that inside the topos $\mathbf {Sets}$ the unique
localization measure of observables is a point of the $\mathcal
R$-spectrum of the corresponding algebra of scalars, which is
assigned a numerical identity. In contradistinction, inside the
topos $\mathbf {Shv}(X)$, the former is substituted by a variety
of localization measures, dependent only on the open sets in the
topology of $X$. In the latter context, a point-localization
measure, is identified precisely with the ultrafilter of all opens
containing the point. This identification permits the conception
of other filters being formed by admissible operations between
opens as generalized measures of localization of observables.
Furthermore, the relativization of representability in $\mathbf
{Shv}(X)$ is physically significant, because the operational
specification of measurement environments exists only locally and
the underlying assumption is that the information gathered about
local observables in different measurement situations can be
collated together by appropriate means, a process that is
precisely formalized by the notion of sheaf. Conclusively, we
assert that, localization schemes referring to observables may not
depend exclusively on the existence of points, and thus, should
not be tautosemous with the practice of conferring a numerical
identity to them. Thus, the relativization of representability
with respect to the internal reals of the topos of sheaves
$\mathbf {Shv}(X)$, amounts to the substitution of point
localization measures, represented numerically, with localization
measures fibering over the base category of open reference loci,
represented respectively by local sections in the sheaf of
internal reals [22].

The main purpose of the discussion of relativized representability
inside the topos $\mathbf {Shv}(X)$, is again to focus attention
in the fact that the purely algebraic functorial dynamical
mechanism of connections, depending only on the algebra theoretic
structure, still holds inside that topos. The interpretation of
the mechanism in $\mathbf {Shv}(X)$ has been accomplished by the
development of Abstract Differential Geometry (ADG) [21, 26-27]. In particular, ADG
generalizes the differential geometric mechanism of smooth
manifolds, by explicitly demonstrating that most of the usual
differential geometric constructions can be carried out by purely
sheaf-theoretic means without any use of any sort of
$C^\infty$-smoothness or any of the conventional calculus that
goes with it. Thus, it permits the legitimate use of any
appropriate $\mathcal {R}$-algebra sheaf of observables localized
over a topological environment, even Rosinger's singular algebra
sheaf of generalized functions, without loosing the differential
mechanism, prior believed to be solely associated with smooth
manifolds [23-29].

The operational machinery of ADG is essentially based on the
exactness of the following abstract de Rham complex, interpreted inside
the topos of sheaves $\mathbf {Shv}(X)$: $$\mathbf 0 \rightarrow \mathcal R \rightarrow \mathcal A \rightarrow {{\mathbf {{\mathbf
{\Omega}}}}^1({\mathcal A})} \rightarrow {\ldots} \rightarrow
{{\mathbf {{\mathbf {\Omega}}}}^n({\mathcal A})} \rightarrow
{\ldots}$$

It is instructive to note that the  exactness
of the complex above, for the classical case, where, $\mathcal A$ stands for 
a smooth $\mathcal {R}$-algebra sheaf of observables on $X$,
constitutes an expression of the lemma of Poincar\'{e}, according
to which, every closed differential form on $X$ is exact at
least locally in $X$. ADG's power of abstracting and generalizing
the classical calculus on smooth manifolds basically lies in the
possibility of assuming other more general coordinates sheaves,
that is, more general commutative coefficient structure sheaves
$\mathcal A$, while at the same time retaining, via the exactness of the
abstract de Rham complex, as above, the mechanism of differentials,
instantiated, in the first place, in the case of classical
differential geometry on smooth manifolds. Thus, any cohomologically
appropriate sheaf of algebras  $\mathcal A$ may be regarded as a coordinates sheaf 
capable of providing a
differential geometric mechanism, independently of any manifold
concept, analogous, however, to the one supported by smooth
manifolds.

Applications of ADG include the reformulation
of Gauge theories in sheaf-theoretic terms [26, 27], as well as,
the evasion of the problem of manifold singularities appearing in
the context of General Relativity [28-31]. Related with the first
issue, ADG has modeled Yang-Mills fields in terms of appropriate
pairs $(\mathbf E, {\mathcal D}_{\mathbf E})$, where $\mathbf E$
are vector sheaves whose sections have been identified with the
states of the corresponding particles, and ${\mathcal D}_{\mathbf
E}$ are connections that act on the corresponding states causing
interactions by means of the respective fields they refer to.
Related with the second issue, ADG has replaced the $\mathcal
R$-algebra $\mathcal {C^\infty}({M})$ of smooth real-valued
functions on a differential manifold with a sheaf of $\mathcal
{R}$-algebras that incorporates the singularities of the manifold
in terms of appropriate ideals, allowing the formulation of
Einstein's equations in a covariant form with respect to the
generalized scalars of that sheaf of $\mathcal {R}$-algebras.

\subsection{Quantization as a Grothendieck Topos-Relativized Representability and 
Quantum Sheafification of Dynamics}

The basic defining feature of Quantum Theory according to the
Bohrian interpretation [32-34], in contradistinction to all
classical theories, is the realization that physical observables
are not definitely or sharply valued as long as a measurement has
not taken place, meaning both, the precise specification of a
concrete experimental context, and also, the registration of a
value of a measured quantity in that context by means of an
apparatus. Furthermore, Heisenberg's uncertainty relations,
determine the limits for simultaneous measurements of certain
pairs of complementary physical properties, like position and
momentum. In a well-defined sense, the uncertainty relations may
be interpreted as measures of the valuation vagueness associated
with the simultaneous determination of all physical observables in
the same experimental context. In all classical theories, the
valuation algebra is fixed once and for all to be the division
algebra of real numbers $\mathcal R$, reflecting the fact that
values admissible as measured results must be real numbers,
irrespective of the measurement context and simultaneously for all
physical observables.

The resolution of valuation vagueness in Quantum Theory can be
algebraically comprehended through the notion of relativization of
representability of the valuation algebra with respect to commutative
algebraic contexts that correspond to locally prepared measurement
environments [35, 36]. Only after such a relativization the
eigenvalue equations formulated in the context of such a
measurement environment yield numbers corresponding to measurement
outcomes. At a logical level commutative contexts of measurement
correspond to Boolean  algebras, identified as subalgebras of a
quantum observables algebra. In the general case, commutative
algebraic contexts are identified with commutative $\mathcal
K$-algebras, where $\mathcal K= \mathcal {Z}_2, \mathcal R,
\mathcal C$, which may be thought as subalgebras of a
non-commutative algebra of quantum observables, represented
irreducibly as an algebra of operators on a Hilbert space of
quantum states.

If we consider the relativization of representability of the
valuation algebra in Quantum Theory seriously, this implies that the
proper topos to apply the functorial dynamical mechanism caused by
quantum fields, is the topos of sheaves of algebras over the base
category of commutative algebraic contexts, denoted by $\mathbf
{Shv}(\mathcal {A_C})$, where $\mathcal {A_C}$ is the base
category of commutative $\mathcal K$-algebras [20, 22, 37-39].
Equivalently, this means that representability in Quantum Theory
should be relativized with respect to the internal reals of the
topos $\mathbf {Shv}(\mathcal {A_C})$. We mention that the interpretational aspects of
the proposed topos-theoretic relativization of physical representability in relation to
the truth-values structures of quantum logics have been discussed extensively in [40].

In the current context of enquiry, this admissible
topos-theoretic framework of representability elaborates the
interpretation of the algebraic functorial mechanism of
connections inside the topos $\mathbf {Shv}(\mathcal {A_C})$, thus
allowing the conception of interactions caused by quantum fields
and, in particular the notion of quantum gravitational field. In
this sense, Quantum Gravity should be properly a theory
constructed inside the topos $\mathbf {Shv}(\mathcal {A_C})$,
formulated by means of adaptation of the functorial mechanism of
infinitesimal scalars extension in the regime of this topos. It is
instructive to explain the meaning of internal reals inside this
topos. In analogy with the case of internal reals inside the topos
$\mathbf {Shv}(X)$, where the valuation algebra of real numbers has
been relativized with respect to the base category of open sets of an
abstract topological space, in the topos $\mathbf {Shv}(\mathcal
{A_C})$ the valuation algebra is relativized  with respect to the
base category of commutative subalgebras of the algebra of quantum
observables. Thus, similarly with the former case, it admits a
description as a sheaf of continuous local sections over the
category of base reference loci of variation $\mathcal {A_C}$. Of
course, in the quantum case the notion of topology with respect to
which continuity may be conceived refers to an appropriate
Grothendieck topology [17, 19, 20, 22], formulated in terms of covers on the
base category $\mathcal {A_C}$, with respect to which sheaves of
algebras may be defined appropriately. A Grothendieck topology
suitable for this purpose can be explicitly constructed as a
covering system $S$ of epimorphic families on the base category of
commutative contexts, defined by the requirement that the morphism $$ G_S
: {\coprod}_{\{s: {{A_C}^{'}} \rightarrow {A_C} \} \in S} {{A_C}^{'}}
\rightarrow {A_C}$$ where, $A_C$, ${{A_C}^{'}}$ in ${\mathcal A}_C$, is an epimorphism in ${\mathcal A}_Q$. This
topological specification incorporates the functorial notion of
generalized localization schemes, appropriately adapted for
probing the quantum regime of observable structure [20, 22, 37].

The research initiative based on the principle of relativized
representability inside the topos $\mathbf {Shv}(\mathcal {A_C})$,
as the proper universe of discourse for constructing a categorical
theory of covariant Quantum Gravitational Dynamics is in the phase
of intense development, while a nucleus of basic ideas, methods and results related
with this program have been already communicated [20]. According
to this schema, the representation of quantum observables algebras
${A_Q}$ in the category ${\mathcal {A_Q}}$ in terms of sheaves
over commutative arithmetics ${{A}_C}$ in ${\mathcal {A_C}}$ for
the Grothendieck topology of epimorphic families on ${\mathcal
{A_C}}$, is based on the existence of the adjunctive
correspondence ${\mathbf L} \dashv {\mathbf F}$ as follows:
$$\mathbf L :  {{\bf Sets}^{{{\mathcal {A_C}}}^{op}}}
\leftrightarrows  { {\mathcal {A_Q}}} : \mathbf F$$ which says
that the Grothendieck functor of points of a quantum observables algebra
restricted to commutative arithmetics defined by: $${\mathbf
F}({{A_Q}}) : {{A_C}} {\mapsto} {{Hom}_{ {\mathcal
{A_Q}}}({\mathbf M}({{A_C}}), {{A_Q}})}$$ has a left adjoint:
$$\mathbf L : {{\bf Sets}^{{{\mathcal {A_C}}}^{op}}} \to {
{\mathcal {A_Q}}}$$ which is defined for each presheaf $\mathbf P$
in ${{\bf Sets}^{{{\mathcal {A_C}}}^{op}}}$ as the colimit:
$${\mathbf {L}}({\mathbf P})= {\mathbf P} {\otimes}_{\mathcal A_C}
{\mathbf M}$$ where ${\mathbf M}$ is a coordinatization functor,
viz.: $${\mathbf M}:{{\mathcal A}_C} \rightarrow { {\mathcal
A}_Q}$$ which assigns to commutative observables algebras in
${{\mathcal A}_C}$ the underlying quantum algebras from ${
{\mathcal A}_Q}$. Equivalently, there exists a bijection, natural
in $\mathbf P$ and ${{A_Q}}$ as follows: $$ Nat({\mathbf
P},{\mathbf F}({{A_Q}})) \cong {{Hom}_{\mathcal {{A_Q}}}({\mathbf
L \mathbf P}, {{A_Q}})}$$ The counit of the adjunction:
$${\epsilon}_{A_Q} : \mathbf L {\mathbf F}({A_Q}) \rightarrow
{A_Q}$$ defined by the composite endofunctor: $$\mathbf G:=\mathbf
L {\mathbf F}: \mathcal {A_Q} \rightarrow  \mathcal {A_Q} $$
constitutes intuitively the first step of a functorial free
resolution of a quantum observables algebra ${A_Q}$ in $\mathcal
{A_Q}$. Actually, by iterating the endofunctor $\mathbf G$, we may
extend ${\epsilon}_{A_Q}$ to a free simplicial resolution of
${A_Q}$. In this setting, we may now apply K\"{a}hler's methodology in
order to obtain the object of quantum differential $1$-forms, by
means of the following split short exact sequence: 
$$0 \rightarrow
{{\Omega}}_{A_Q} \rightarrow {\mathbf G}{{A_Q}} \rightarrow {A_Q} $$
or equivalently, $$0 \rightarrow
{{\Omega}}_{A_Q} \rightarrow {\mathbf F}({A_{Q}})
{\otimes}_{\mathcal A_C} {\mathbf M} \rightarrow {A_Q} $$
According to the above, we obtain that: $$
{{\Omega}_{A_Q}}=\frac{J}{{J}^2}$$ where ${J}= {\mathbf
{Ker}}({\epsilon}_{A_Q})$ denotes the kernel of the counit of the
adjunction. Subsequently, we may apply the algebraic construction,
for each $n \in N$, $n \geq 2$, of the $n$-fold exterior product
${{{{\Omega}^n}_{A_Q}}}={{\bigwedge}^n} {{{{\Omega}^1}_{A_Q}}}$.
Thus, we may now set up the algebraic de Rham complex of ${A_Q}$
as follows: $${A_Q} \rightarrow {{\Omega}_{A_Q}} \rightarrow
{\ldots} \rightarrow {{{{\Omega}^n}_{A_Q}}} \rightarrow {\ldots}$$
At a next stage, we notice that the functor of points of a quantum
observables algebra restricted to commutative arithmetics
${\mathbf F}({{A_Q}})$ is left exact, because it is the right
adjoint functor of the established adjunction. Thus, it preserves
the short exact sequence defining the object of quantum
differential $1$-forms, in the following form: $$0 \rightarrow
{\mathbf F} ({{\Omega}}_{A_Q}) \rightarrow {\mathbf F} ({\mathbf
G}({A_{Q}})) \rightarrow {\mathbf F}({{A_Q}})$$ Hence, we
immediately obtain that: $${\mathbf F} ({{\Omega}}_{A_Q})=
\frac{Z}{{Z}^2}$$ where ${Z}= {\mathbf {Ker}}({\mathbf F} (
{\epsilon}_{A_Q}))$. Then, in analogy with the general algebraic
situation, interpreted inside the proper universe that the functor
of points of a quantum observables algebra assumes existence,
viz., the topos ${{\bf Sets}^{{{\mathcal {A_C}}}^{op}}}$, we
introduce the notion of an interaction field, termed quantum
field, by means of the functorial pair $\big ({\mathbf
F}({{A_Q}}):= {{Hom}_{ {\mathcal {A_Q}}}({\mathbf M}({{-}}),
{{A_Q}})}, {{\mathcal {{\nabla}}}}_{{\mathbf F}({{A_Q}})} \big)$,
where the quantum connection ${{\mathcal {{\nabla}}}}_{{\mathbf
F}({{A_Q}})}$ is defined as the following natural transformation:
$${{\mathcal {{\nabla}}}}_{{\mathbf F}({{A_Q}})}: {\mathbf F}
({A_Q}) \rightarrow {\mathbf F} ({{\Omega}}_{A_Q})$$ Thus, the
quantum connection ${{\mathcal {{\nabla}}}}_{{\mathbf
F}({{A_Q}})}$ induces a sequence of functorial morphisms, or
equivalently, natural transformations as follows: $${{\mathbf
F}({{A_Q}})} \rightarrow {\mathbf F} ({{\Omega}}_{A_Q})
\rightarrow {\ldots} \rightarrow {\mathbf F}
({{{\Omega}}^n}_{A_Q}) \rightarrow {\ldots}$$ Let us denote by:
$${\mathbf R}_\nabla : {{\mathbf F}({{A_Q}})} \rightarrow {\mathbf
F} ({{{\Omega}}^2}_{A_Q})$$ the composition ${{\mathcal
{{\nabla}}}}^1 \circ {{\mathcal {{\nabla}}}}^0$ in the obvious
notation, where ${{\mathcal {{\nabla}}}}^0:={{\mathcal
{{\nabla}}}}_{{\mathbf F}({{A_Q}})}$, which we call the curvature
of the quantum connection ${{\mathcal {{\nabla}}}}_{{\mathbf
F}({{A_Q}})}$. The latter sequence of functorial morphisms, is
actually a complex if and only if ${\mathbf R}_\nabla=0$. We say
that the quantum connection ${{\mathcal {{\nabla}}}}_{{\mathbf
F}({{A_Q}})}$ is integrable or flat if ${\mathbf R}_\nabla=0$,
referring to the above complex as the functorial de Rham complex
of the integrable connection ${{\mathcal {{\nabla}}}}_{{\mathbf
F}({{A_Q}})}$ in that case. The vanishing of the curvature of the
quantum connection, that is: $${\mathbf R}_\nabla=0$$  can be
used as a means of transcription of Einstein's equations in the
quantum regime, that is inside the topos $\mathbf {Shv}(\mathcal
{A_C})$ of sheaves of algebras over the base category of
commutative algebraic contexts, in the absence of cohomological
obstructions. We may explain the curvature of the quantum
connection as the effect of non-trivial interlocking of
commutative arithmetics, in some underlying diagram of a quantum
observables algebras being formed by such localizing commutative
arithmetics. The non-trivial gluing of commutative arithmetics in
localization systems of a quantum algebra is caused by topological
obstructions, that in turn, are being co-implied by acyclicity of
the algebraic de Rham complex of ${A_Q}$. Intuitively, a
non-vanishing curvature is the non-local attribute detected by an
observer employing a commutative arithmetic in a discretely
topologized categorical environment, in the attempt to understand
the quantum localization properties, after having introduced a
potential (quantum gravitational connection) in order to account
for the latter by means of a differential geometric mechanism [20].
Thus, the physical meaning of curvature is associated with the
apparent existence of non-local correlations from the restricted
spatial perspective of disjoint classical commutative arithmetics
$A_{C}$. It is instructive to make clear that, in the present
schema, the notion of curvature does not refer to an underlying
background manifold, since such a structure has not been required
at all in the development of the differential geometric mechanism,
according to functorial homological algebraic methods.

\section{Epilogue}

Conclusively, it is worthwhile to emphasize that discussions of
background manifold independence pertaining the current research
focus in Quantum Gravity, should take at face value the fact that
the fixed manifold construct in General Relativity is just the
byproduct of fixing physical representabilty in terms of real
numbers. Moreover, it is completely independent of the possibility
of formulating dynamics, since the latter can be developed
precisely along purely algebraic lines, that is, by means of
functorial connections. Hence the usual analytic differential
geometric framework of smooth manifolds, needed for the
formulation of General Relativity, is just a special
coordinatization of the universal functorial mechanism of
infinitesimal scalars extension, and thus should be substituted
appropriately, in case a merging with Quantum Theory is being
sought. The substitution is guided by the principle of relativized
representability with respect to a suitable topos.

The important physical issue incorporated in the idea of relativizing physical representability with respect to appropriate topos-theoretic universes of discourse, as has been explained previously, has to do with a novel conception of physical localization schemes, that, in particular, seem to be indispensable for an accurate comprehension of the quantum regime of observable structure. More concretely, in classical theories localization has been conceived by means of metrical properties on a pre-existing smooth set-theoretic spacetime manifold. In contradistinction, quantum localization should be understood categorically and algebraically, viz., purely in functorial terms of relational information content between quantum arithmetics (algebras of quantum observables) and diagrams of commutative ones, without any supporting notion of smooth metrical backround manifold. In this sense, the resolution focus should be shifted from point-set to topological localization models of
quantum observables structures, that effectively, induce a
transition in the semantics of observables from a set-theoretic to
a sheaf-theoretic one. Subsequently, that semantic transition effectuates the conceptual replacement of the classical metrical ruler of localization on a smooth background manifold, with a sheaf-cohomological ruler of algebraic categorical localization in a Grothendieck topos, that captures the relational information of observables in the quantum regime, filtered through diagrams of local commutative arithmetics (local reference frames in the topos constituting the functor of points of a quantum arithmetic). Thus, the dynamical properies of quantum arithmetics are being properly addressed to the global topos-theoretic dynamics generated by categorical diagrams of local commutative ones, giving rise to complexes of sheaves of observables for a suitably defined Grothendieck topology consisting of epimorphic families of coverings by local commutative arithmetics.

In particular, the application of the principle of relativized representability on the problem of 
merging General Relativity with Quantum Theory, forces the topos of sheaves of commutative
observables algebras $\mathbf {Shv}(\mathcal {A_C})$ as the proper
universe of discourse for Quantum Gravity, requiring a functorial
adaptation of the algebraic mechanism of connections inside that
topos, and subsequently, an interpretation of Quantum Gravitational
dynamics sheaf cohomologically with respect to the non-trivial
localization schemes of observables in the quantum regime.


\vspace{10mm}
\begin{thebibliography}{99}



\bibitem [1]. A. Ashtekar,  and J. Lewandowski, {\it Background independent quantum gravity: a status report},
pre-print (2004); gr-qc/0404018.


\bibitem [2]. J. Butterfield, and C. J. Isham,
{\it Some Possible Roles for Topos Theory in Quantum Theory and
Quantum Gravity}, Foundations of Physics, {\bf 30}, 1707 (2000).

\bibitem [3]. L. Crane, {\it Clock and Category: Is Quantum Gravity
Algebraic?}, Journal of Mathematical Physics, {\bf 36}, 6180
(1995).

\bibitem[4]. D. R. Finkelstein,
{\it Quantum Relativity: A Synthesis of the Ideas of Einstein and
Heisenberg}, Springer-Verlag, Berlin-Heidelberg-New York (1996).


\bibitem[5]. S. A. Selesnick,
{\it Quanta, Logic and Spacetime}, (2nd. ed), World Scientific
(2004).


\bibitem [6]. C. J. Isham, {\it Some Reflections on the Status of Conventional Quantum Theory when Applied to
Quantum Gravity}, in {\itshape The Future of Theoretical Physics
and Cosmology: Celebrating Stephen Hawking's 60th Birthday},
Gibbons, G. W., Shellard, E. P. S. and Rankin, S. J. (Eds.),
Cambridge University Press, Cambridge (2003); quant-ph/0206090.

\bibitem [7]. C. J. Isham, {\it A new approach to quantising space-time:
I. Quantising on a general category}, Advances in Theoretical and
Mathematical Physics, {\bf 7}, 331 (2003); gr-qc/0303060.



\bibitem [8]. R. Penrose, {\it The problem of spacetime singularities: implications for quantum
gravity?}, in {\itshape The Future of Theoretical Physics and
Cosmology: Celebrating Stephen Hawking's 60th Birthday}, Gibbons,
G. W., Shellard, E. P. S. and Rankin, S. J. (Eds.), Cambridge
University Press, Cambridge (2003).


\bibitem [9]. L. Smolin, {\it The case for background independence}, pre-print (2005); gr-qc/0507235.


\bibitem [10]. R. D. Sorkin,
{\it A Specimen of Theory Construction from Quantum Gravity}, in
{\it The Creation of Ideas in Physics}, Leplin, J. (Ed.), Kluwer
Academic Publishers, Dordrecht (1995); gr-qc/9511063.

\bibitem [11]. J. Stachel, {\it Einstein and Quantum Mechanics}, in
{\it Conceptual Problems of Quantum Gravity}, Ashtekar, A. and
Stachel, J. (Eds.), Birkh\"auser, Boston-Basel-Berlin (1991).



\bibitem [12]. M. Atiyah and I. MacDonald, {\it Introduction to Commutative Algebra}, AddisonWesley, Reading, Massachusetts, 1969.


\bibitem [13]. S. MacLane, {\it Categories for the Working
Mathematician}, Springer-Verlag, New York, 1971.


\bibitem [14]. I. R. Shafarevich, {\it Basic Notions of Algebra},
Springer, Berlin (1997).






\bibitem [15]. J. L. Bell,  {\it From Absolute to Local Mathematics},
Synthese {\bf 69}, 1986.


\bibitem [16]. J. L. Bell,  {\it Categories, Toposes and Sets},
Synthese, {\bf 51}, No.3, 1982.




\bibitem [17]. S. MacLane and I. Moerdijk, {\it Sheaves in Geometry
and Logic}, Springer-Verlag, New York, 1992.


\bibitem [18]. J. L. Bell, {\it Toposes and Local Set Theories},
Oxford University Press, Oxford, 1988.






\bibitem [19]. M. Artin, A. Grothendieck, and  J. L. Verdier, {\it
Theorie de topos et cohomologie etale des schemas}, (Springer LNM
269 and 270, Springer-Verlag, Berlin, 1972).


\bibitem [20]. E. Zafiris, {\it Quantum Observables Algebras and Abstract
Differential Geometry: The Topos-Theoretic Dynamics of Diagrams 
of Commutative Algebraic Localizations}, International Journal of Theoretical
Physics {\bf 46}, (2) (2007).






\bibitem
[21]. A. Mallios, {\it Geometry of Vector Sheaves: An Axiomatic
Approach to Differential Geometry}, Vols 1-2, Kluwer Academic
Publishers, Dordrecht (1998).


\bibitem [22]. E. Zafiris, {\it Generalized Topological Covering
Systems on Quantum Events Structures}, Journal of Physics A:
Mathematical and General {\bf 39}, (2006).



\bibitem [23]. A. Mallios, {\it Remarks on
``singularities''}, Progress in Mathematical Physics, Columbus, F.
(Ed.), Nova Science Publishers, Hauppauge, New York (2003)
(invited paper); gr-qc/0202028.

\bibitem [24].  A. Mallios, {\it Quantum gravity and ``singularities''},
Note di Matematica, {\bf 25}, 57 (2005)/(2006) (invited paper);
physics/0405111.

\bibitem [25].  A. Mallios, {\it Geometry and physics
today}, International Journal of Theoretical
Physics (online first DOI: 10.1007/s10773-006-9130-3) (2006).

\bibitem [26].  A. Mallios, {\it Modern Differential Geometry in
Gauge Theories}: vol. 1. {\it Maxwell Fields}, Birkh\"auser,
Boston, 2006.











\bibitem [27].  A. Mallios, {\it Modern Differential Geometry in
Gauge Theories}:  vol. 2. {\it Yang-Mills Fields}, (forthcoming by
Birkh\"auser, Boston, 2007).


\bibitem
[28]. A. Mallios, E. E. Rosinger, {\it Abstract Differential
Geometry, differential algebras of generalized functions, and de
Rham cohomology},  Acta Appl. Math. {\bf 55}, 231 (1999).



\bibitem
[29]. A. Mallios, E. E. Rosinger, {\it Space-Time foam dense
singularities and de Rham cohomology}, Acta Appl. Math. {\bf 67},
59 (2001).






\bibitem [30].  A. Mallios, and I. Raptis, {\itshape Finitary \v{C}ech-de Rham
Cohomology: much ado without $\mathcal {C^\infty}$-smoothness},
International Journal of Theoretical Physics, {\bf 41}, 1857
(2002); gr-qc/0110033.


\bibitem [31].  A. Mallios, and I. Raptis, {\it $\mathcal {C^\infty}$-Smooth Singularities Exposed:
Chimeras of the Differential Spacetime Manifold}; gr-qc/0411121.



\bibitem [32]. N. Bohr  {\it Atomic Physics and Human Knowledge},
John Wiley, New York, 1958.


\bibitem [33]. H. J. Folse {\it The Philosophy of Niels Bohr. The
Framework of Complementarity}, New York, North Holland, 1985.



\bibitem [34]. J. Bub, {\it Interpreting the Quantum World},
Cambridge University Press, Cambridge, 1997.



\bibitem [35]. E. Zafiris, {\it Probing Quantum Structure Through Boolean
Localization Systems},
International Journal of Theoretical Physics  {\bf 39}, (12)
(2000).

\bibitem [36]. E. Zafiris, {\it Boolean Coverings of Quantum Observable
Structure: A Setting for an Abstract Differential Geometric
Mechanism }, Journal of Geometry and Physics {\bf 50}, 99 (2004).


\bibitem [37]. E. Zafiris, {\it Quantum Event Structures from the perspective of
Grothendieck Topoi},  Foundations of Physics {\bf 34}, (7) (2004).


\bibitem [38]. E. Zafiris, {\it Interpreting Observables in a Quantum World from
the Categorial Standpoint}, International Journal of Theoretical
Physics  {\bf 43}, (1) (2004).


\bibitem [39]. E. Zafiris, {\it  Topos Theoretical Reference Frames on the Category of 
Quantum Observables}, quant-ph/0202057.

\bibitem [40]. E. Zafiris, {\it  Categorical Foundations of Quantum Logics and 
Their Truth Values Structures}, quant-ph/0402174.


\end{thebibliography}
\end{document}